\documentclass[letterpaper]{article}
\usepackage{aaai}

\usepackage{times}
\usepackage{helvet}
\usepackage{courier}
\usepackage{amsmath}
\usepackage{times,graphicx,xspace,comment,subfigure,algorithm,algpseudocode,multirow,url,amsmath}
\begin{document}
\title{Learning the Latent State Space of Time-Varying Graphs}
\author{Nesreen K. Ahmed, Christopher Cole, Jennifer Neville\\
Department of Computer Science, Purdue University\\
West Lafayette, IN 47907\\
\{nkahmed,cole14,neville\}@cs.purdue.edu\\
}
\vspace{-5.mm}
\maketitle
\begin{abstract}
\begin{quote}
From social networks to Internet applications, a wide variety of electronic communication tools are producing streams of graph data---
where the nodes represent users and the edges represent the contacts between them over time. This has led to an increased interest in mechanisms to model the dynamic structure of time-varying graphs. In this work, we develop a framework for learning the latent state space of a time-varying email graph. We show how the framework can be used to find subsequences that correspond to global real-time events in the Email graph (e.g. vacations, breaks, etc.). These events impact the underlying graph process to make its characteristics non-stationary. Within the framework, we compare two different representations of the temporal relationships---discrete vs. probabilistic. We use the two representations as inputs to a mixture model to learn the latent state transitions that correspond to important changes in the Email graph structure over time. 
\end{quote}
\end{abstract}

\section{Introduction}

Time-varying graphs have been extensively studied from the perspective of time-aggregated communications between nodes during a particular time interval~\cite{leskovec2005graphs}. Using this representation, researchers have discovered structural patterns in graphs with long-lived links among the nodes (e.g., hub nodes in the Web~\cite{barabasi1999emergence}). However, today, a wide variety of electronic and online communication tools are  producing streams of graph data. These streams are usually rapid, time-varying, unbounded sequences of short-lived  links among the nodes (e.g., email, SMS, tweets). Since the interactions are indicators of hidden relationship structures among the individuals, there has recently been an interest in moving beyond the perspective of time-aggregated graphs to  model and mine the dynamic structure of graph streams.   

The work in \cite{perra2012activity} highlights the biases that result if time-aggregated representations are used to analyze dynamical processes such as information/disease spread, and discuss the importance of distinguishing between the structural evolution and the dynamical process unfolding on top of its structure. 
Also, the work in \cite{gautreau2009microdynamics} compared the statistical distributions of structural properties across separate, daily snapshots of a US airport network graph and showed that even when the statistical distributions are stationary, there is often intense activity within each graph snapshot, with many links disappearing/appearing. 

In this work, we develop a framework for learning the higher-level latent state space of time-varying graphs. We study data collected from the email logs of university mail servers. The links correspond to the set of email communications among the students/staff/faculty in Purdue university from August $2011$ to February $2012$. This seven month period includes several calendar events (breaks, vacations, ...etc.) that take place during the academic year. These real-time events usually impact the underlying graph process to make its characteristics non-stationary and drift from the overall mean. We show how our framework discovers subsequences that correspond to these global events without detecting local changes that are due to diurnal patterns. 

\section{Framework}
Assume that we have a stream of edges continuously evolving over time $E=\{e_t=(u,v), \forall t | u,v \in V \}$. These edges correspond to the communications among a set of nodes $V$. Here, we consider the time-varying graph $G$ as a sequence of graphs at each timestep, denoted by $\mathcal{G}=\{\mathcal{G}_1,...,\mathcal{G}_t,...\}$. 

\subsection{Models of Representation}
Since we cannot know the real lifetime of the relationships among the nodes in real graphs, we compare two different models for the representation of the relationship lifetime:
\begin{enumerate}
\item \textbf{Discrete Model:} In this representation, edges are aggregated in non-overlapping discrete windows of size $\delta t$, such that any edge $e_{t'}=(u_i,v_j)$ that occurs at time $t'$  will be considered in the graph $\mathcal{G}_t$ if and only if $t' \in [t,t+\delta t]$. We consider $\delta t$ equals $1$ day.

\item \textbf{Probabilistic Model:} In this representation, we consider the graph $G_t$ at time $t$, as a probabilistic graph that contains any edge $e_{t'}=(u_i,v_j)$ that occurs at any time $t'$ such that $t' \leq t$, and $p_{ij}=\exp({-(t-t')/\tau})$ is the probability of the edge between node $u_i$ and node $v_j$, and $\tau$ is the mean lifetime of the edge. These probabilities change over time, and we age-out older edges with probabilities less than a very small cut-off threshold. In this paper, we consider $\tau$ equals $12$ days. 
\end{enumerate}

\noindent
Previous work considered different variation of these models for social network analysis, see~\cite{goyal2010learning,caceres2011temporal} for details.
\subsection{State-Space Model}

\begin{figure}[!h]
\centering
\vspace{-2.mm}
\subfigure[Disc. Model $\delta t=1$]{\label{fig:disc_avgdeg}\includegraphics[width=0.23\textwidth]{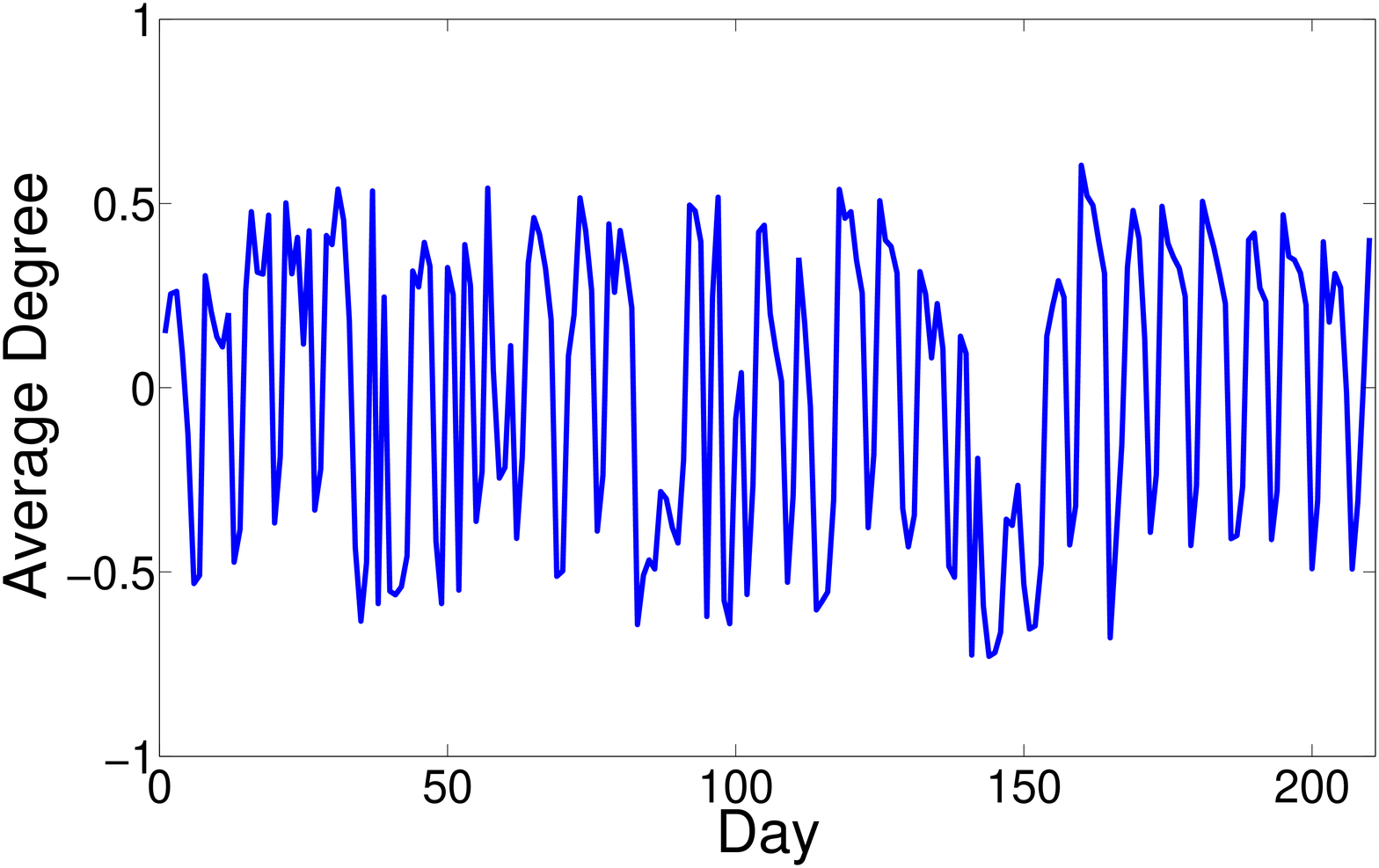}}
\hspace{-1.mm}
\subfigure[Prob. Model $\tau=12$]{\label{fig:prob_avgdeg}\includegraphics[width=0.23\textwidth]{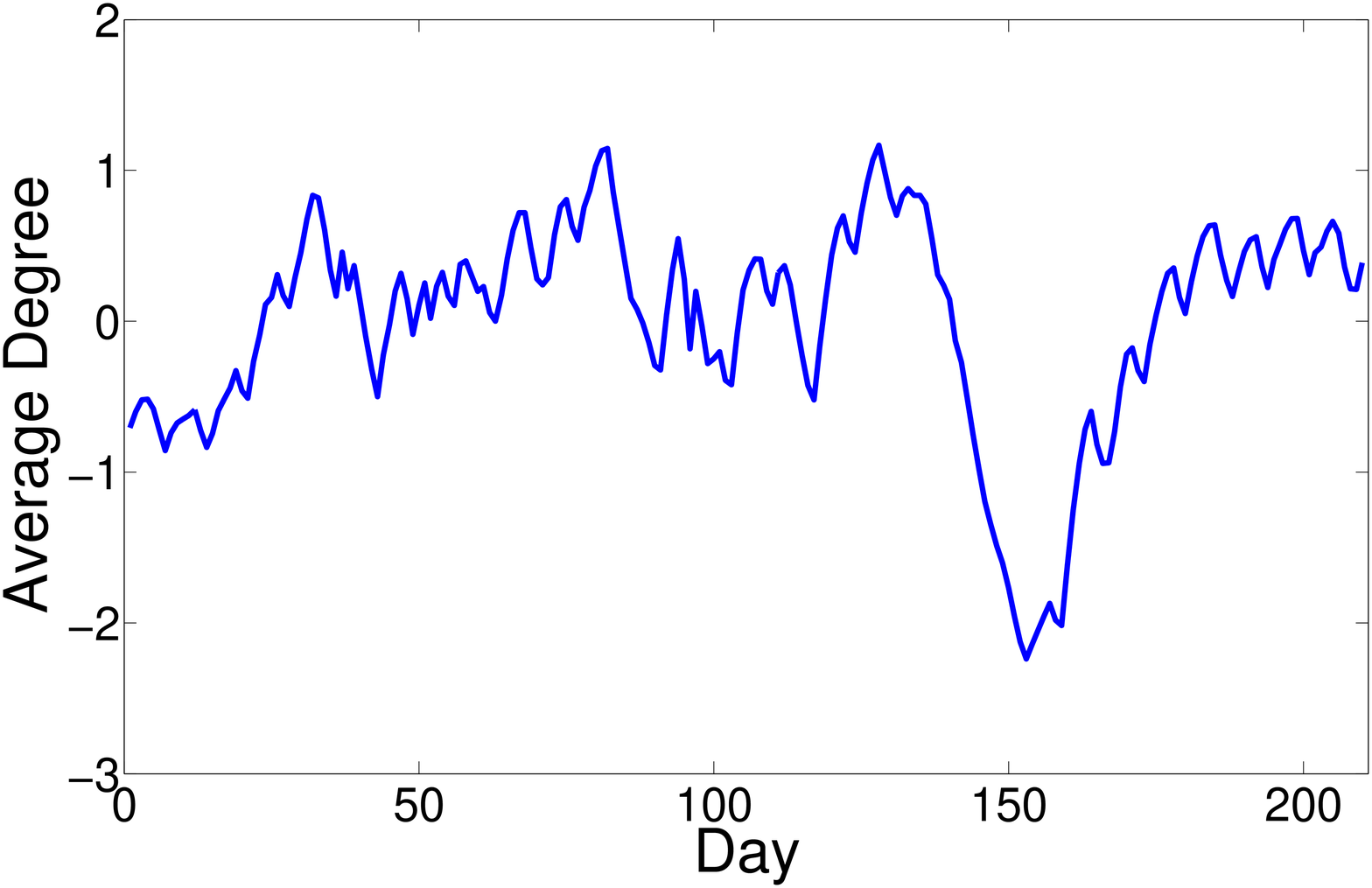}}
\vspace{-3.mm}
\caption{Average degree versus time}
\label{fig:avgdeg}
\vspace{-4.mm}
\end{figure}

The normal operation of any system or dynamical process can be characterized in different temporal states. To explore the state space of a time-varying graph, we represent each graph $\mathcal{G}_t$ in the sequence of graphs $\mathcal{G}$ by a set of attributes $\mathcal{A}$. These attributes correspond to structural properties of each graph $\mathcal{G}_t$. Here, we consider the average degree and average clustering of the graph as attributes. Then, we use the KMeans algorithm to cluster the graphs in $\mathcal{G}$ using their attributes (average degree and average clustering). Each cluster includes all the graphs that have similar properties. Therefore, each cluster represents a state and the state transition diagram corresponds to the cluster memberships of graphs in $\mathcal{G}$ versus time. We select the number of clusters equals seven (i.e., corresponding to seven days).  Figure~\ref{fig:avgdeg} shows the plots of the average degree at each day after subtracting the linear trend to make the data changing around the zero mean (i.e., detrending the timeseries). Clearly, the discrete model shows the local daily changes that take place in $G$, however, the probabilistic model emphasizes the regions in the stream that correspond to global events taking place (as we see next in Results section).

\begin{figure}[!h]
\centering
\vspace{-4.mm}
\subfigure[Disc. Model $\delta t=1$]{\label{fig:disc_state_plot}\includegraphics[width=0.35\textwidth]{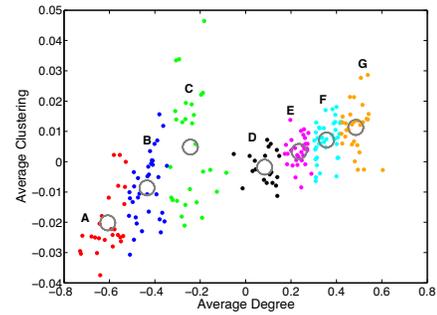}}
\hspace{-7.mm}
\subfigure[Prob. Model $\tau=12$]{\label{fig:prob_state_plot}\includegraphics[width=0.35\textwidth]{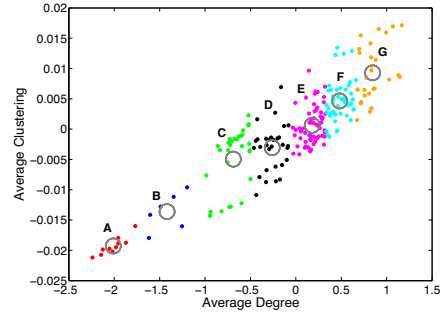}}
\vspace{-4.mm}
\caption{Plot of time-varying graphs as points in the space}
\label{fig:state_plot}
\end{figure}

\begin{figure}[!h]
\centering
\vspace{-4.mm}
\subfigure[Disc. Model $\delta t=1$]{\label{fig:disc_state_trans}\includegraphics[width=0.39\textwidth]{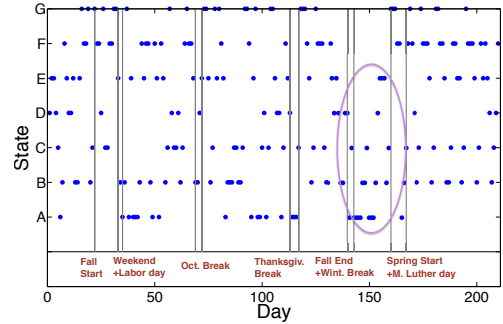}}
\hspace{-7.mm}
\subfigure[Prob. Model $\tau=12$]{\label{fig:prob_state_trans}\includegraphics[width=0.39\textwidth]{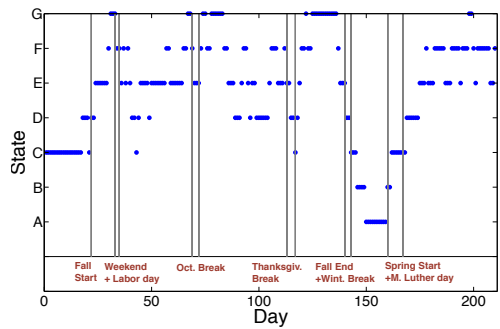}}
\vspace{-4.mm}
\caption{State transition diagram}
\label{fig:state_trans}
\vspace{-4mm}
\end{figure}

\section{Results and Discussion}
Figures~\ref{fig:prob_state_plot}, and~\ref{fig:disc_state_plot} show the scatter plots of graphs in the sequence $\mathcal{G}$ as points in the space colored by their state. Also, Figures~\ref{fig:prob_state_trans}, and~\ref{fig:disc_state_trans} show the state transition plot versus time for both the two models. Note that we clustered the graphs into seven states from A to G. Overall, the results indicate that the probabilistic model is more capable of tracking the global events taking place during the fall semester. For example, consider the subsequence between the events of "Wint. Break" (corresponding to the start of the winter break) until "Spring Start" (corresponding to the start of the Spring semester), the probabilistic model detects the break event as a subsequence of very low activity graphs (state A), when most of the students will be out of campus. However, the discrete model shows local email communications that possibly correspond to staff members, missing the global break event. In addition, the probabilistic model can distinguish between different types of events and breaks based on the intensity of their effect on the graph structure. For instance, the "Thanksgiving Break" and the "Fall Break" are labeled as two different states C and A. However, the discrete model treats them as state A. 

To summarize, in this work, we developed a framework for learning the higher-level latent state space of time-varying graphs. The proposed framework is based on the probablistic representation model of graph streams. In future work, we aim to study the problems of mining and modeling probabilistic graph streams and how to use these models to predict the global characteristics of the graph strcuture in the next timesteps.

\bibliographystyle{aaai}
\bibliography{paper}

\end{document}